\documentclass{iopart}
\usepackage{iopams}

\usepackage{amsfonts,amssymb,bbm,citesort,graphicx}

\newcommand{\eqn}[1]{\begin{equation} #1 \end{equation}} 
\newcommand{\mc}{\mathcal}                               
\newcommand{\mbf}{\mathbf}                               
\newcommand{\mrm}{\mathrm}                    
\newcommand{\1}{\mathbbm 1}                  
\newcommand{\eq}[1]{(\ref{#1})}              
\newcommand{\pd}{\partial}                   
\newcommand{\wtilde}{\widetilde}             
\newcommand{\wbar}{\overline}                
\renewcommand{\l}{\left}                     
\renewcommand{\r}{\right}                    

\begin{document}

\review[Multimode random lasers]{Recent developments in the theory of multimode
random lasers}

\author{Oleg Zaitsev}
\address{Fachbereich Physik, Universit\"at Duisburg-Essen,
             Lotharstr.~1, 47048 Duisburg, Germany}
\ead{oleg.zaitsev@uni-duisburg-essen.de}

\author{Lev Deych}
\address{Physics Department, Queens College of City University of
             New York, Flushing, NY 11367, U.S.A.}
\ead{lev.deych@qc.cuny.edu}

\begin{abstract}

We review recent extensions of semiclassical multimode laser theory to open
systems with overlapping resonances and inhomogeneous refractive index. An
essential ingredient of the theory are biorthogonal quasimodes that describe
field decay in an open passive system and are used as a basis for lasing modes.
We discuss applications of the semiclassical theory, as well as other
experimental and numerical results related to random lasing with mode
competition.

\end{abstract}


\section{Introduction}\label{sec:intro}

The term ``random lasing'' encompasses a number of phenomena related
to light amplification in systems characterized by  a spatial
distribution of electromagnetic field which is much more complex
(irregular) than well-defined cavity modes of standard lasing
structures. Most directly this term describes emission of light by
spatially inhomogeneous disordered materials not bounded by any
artificial mirrors, even though it is also used in the case of
systems with well-defined cavities, which, however, are
characterized by chaotic ray dynamics. These two classes of random
lasers are significantly different, but in some situations, when the
statistical properties of the field in disordered system and in
chaotic cavities are similar, the emission properties of both lasers
can be discussed on equal footing~\cite{misi98}. In the course of
the last decade, random lasing has been observed in many different
types of disordered materials (polymer films~\cite{pols01b}, porous
materials~\cite{mole07}, powders~\cite{cao99},
ceramics~\cite{baho02,baho03},
clusters~\cite{cao00}, colloidal solutions of
nanoparticles~\cite{wu06}), so that it can be regarded as a universal
property of disordered structures.

Lasing in any system is produced  due to combination of two factors:
optical amplification and feedback. In regular lasers the presence of a
feedback always assumes existence of well-defined phase relations between waves
propagating in opposite directions. In a different language the presence of a
feedback is described as existence of well-defined long-living cavity modes
characterized by a regular spatial pattern of electromagnetic field, which sets
inside the lasing structure in the stationary regime.
In~\cite{leto67a,leto67b}, where the concept of random lasers was conceived, a
possibility of lasing caused by a different type of a feedback was proposed. It
was shown that even if propagation of light is described in terms of diffusion
equation, which completely ignores its wave nature and does not have any phase
information, one can still have a laser-like behavior of emission characterized
by a threshold, spectral narrowing, relaxation oscillations and other
attributes of laser oscillations. Physical origin of this phenomenon lies in a
significant increase of a length of light trajectory inside finite amplifying
volume due to multiple scattering. Since amplification results in exponential
growth of light intensity with the distance traveled inside the gain medium, it
can be characterized by gain length $L_g$, which depends on the gain factor and
the diffusion coefficient of light in the medium. The transition to lasing
occurs when the gain length exceeds the loss length, $L_l$, so that the
threshold condition can be written down as $L_g=L_l$. This type of feedback is
called incoherent or nonresonant feedback. The latter term refers to the
absence of any resonant features in the distribution of the field inside the
gain medium or, in other words, the absence of any distinct modes: the spatial
distribution of light intensity is the same regardless of its frequency. In
this situation the lasing frequency is determined by only one remaining
resonant element of the system---the atomic transition, so that the single-peak
emission spectrum is the characteristic feature of lasing with nonresonant
feedback~\cite{leto67a,leto67b}. The authors of~\cite{amba66a,amba66b}, where
non-resonant feedback was discussed in connection with lasing in cavities with
rough surfaces, formulated four conditions required to realize such a complete
nonresonant situations: (i) no mode degeneracy, (ii) equal mode loss, (iii)
mode overlapping (the resonance width of each mode due to radiative losses must
be larger than mean intermode spacing) and (iv) mode mixing (there must be
processes causing frequency of emitted photons to change by an amount larger
than the spectral distance between the modes). While conditions (i), (iii) and
(iv) are realized automatically in disordered systems of large enough
dimensions and chaotic cavities, the condition (ii) is more difficult to
fulfill. Indeed, distribution of widths of scattering resonances in chaotic
cavities and disordered systems was found to be rather broad with different
modes having decay rates differing by the orders of
magnitude~\cite{kott05,kott05}. However, when interest to random lasers was
renewed in 1994 following the observation of lasing from solution of dye
molecules surrounded by titanium dioxide particles~\cite{lawa94}, the
importance of this condition was apparently underappreciated. Namely, results
of~\cite{lawa94} and subsequent experiments were usually explained on the basis
of the concept of the nonresonant feedback with its underlying assumption that
the transport of light can be described within diffusion approximation (see,
e.g.,~\cite{john96,wier96,john04,john04b,pier07}). Validity of diffusion
approximation is usually associated with weak scattering of light, when the
system is far away from Anderson localization transition and interference
effects are weak. Therefore, nonresonant or incoherent feedback was by
extension associated with the regime of weak scattering of light (see,
e.g.,~\cite{cao03c,cao03}).

Inducing stronger light scattering by increasing concentration of
scatterers, it became possible to observe a qualitatively new
phenomenon~\cite{cao99,frol99b,cao00b,ling01}: with
increasing scattering multiple emission lines appeared in the
spectrum, instead of just a single peak at weaker scattering. New
peaks, characterized by much narrower line widths, emerged one by
one with increasing pumping. Since incoherent feedback can only
result in a single-frequency emission spectrum, it was suggested in
\cite{cao99} that the changes in the emission spectrum are due to
transition from incoherent to the coherent feedback. This idea was
supported by studies of photon statistics, which showed that, similar
to regular lasers, light emitted at the peak frequencies had Poisson
photon count distribution~\cite{cao01,pols01b}.

Originally it was suggested in~\cite{cao99} that the feedback is provided by
randomly occurring closed trajectories formed by multiply scattered light.
Eventually this idea was developed to a more general concept of random cavities
(resonators), arising in strongly scattering medium. Anderson localization was
considered as one of the possible mechanisms responsible for formation of such
cavities~\cite{cao00c,vann01,jian02}. In order to verify this assumption a
great deal of efforts have been devoted to analyzing lasing in one-dimensional
models in which all states are
localized~\cite{buri02,souk02,li01,cao03b,miln05}. The localization-based
approach allowed to explain a number of experimentally observed results such as
mode repulsion and saturation of the number of lasing modes~\cite{jian00}.
However, it has never been convincingly demonstrated that light in strongly
scattering three- or two-dimensional samples was indeed close to the Anderson
transition. An alternative mechanism of formation of cavities that could be
responsible for coherent feedback in random media was put forward
in~\cite{apal02}, where it was suggested that random fluctuations of the
refractive index of a disordered medium can result in macroscopically large
ring-like configurations capable of trapping light for long times and serve,
therefore, as random resonators. This model was supported by studies of lasing
in $\pi$-conjugated polymers~\cite{pols02,pols03,pols05}, where certain degree
of universality in spectral distribution of lasing modes was found. This
universality was explained by noting that among multiple random resonators only
those with largest Q-factors (optimal resonators) mostly contribute to lasing.
While distribution of all resonators is very broad, the optimal resonators have
almost identical characteristics, thus explaining observed universality. These
random light-trapping configurations are analogous to so called prelocalized
states, known to exist in the case of electrons in random potentials, and can
arise even when a disordered system is far away from the localization
transition. It was, however, shown in~\cite{apal02} that the trapping
configurations can appear with any appreciable probability only if spatial
fluctuations of the refractive index are correlated over large enough
distances, which might be a reasonable assumption for the polymer samples
studied in~\cite{pols02,pols03,pols05}, but is more difficult to justify for
the ZnO powders~\cite{cao99}. A yet another alternative model of random lasing
was proposed in~\cite{herr98,wilh98}, where random lasers were treated as
lasers with distributed feedback. This approach is somewhat similar to the
random-resonator model with only difference that, instead of dealing with
ring-like resonators, the distributed-feedback model assumed that long-scale
almost periodic Bragg-like configurations are responsible for lasing. This
model, however, has not yet been sufficiently developed.

More recent developments in the field of random lasers have been associated
with renewed attention to the weakly scattering samples. Emission spectrum
containing very sharp spikes was observed in very weakly scattering samples
in~\cite{muju04}. It was argued in that paper that the appearance of the
observed peaks did not require any feedback and could be explained by assuming
that some of spontaneously emitted ``photons'' travel much-longer-than-average
distances. This idea was supported by simulations based on random walk model
(hence the quotation marks in the ``photon''), which reproduced fairly well the
experimental data. However, it was found in~\cite{wu07,wu08} that there are two
types of lines in the emission spectrum of a random laser. One, called
``spikes'' in \cite{wu07,wu08}, similar to that observed in~\cite{muju04}, was
detected (with strong enough pumping) even in samples without any scatterers at
all. The second type, called ``peaks'', only appeared in the presence of
scatterers. The spikes and peaks demonstrated significantly different
statistical properties, which allowed the authors of~\cite{wu07,wu08} to
attribute the origin of the former to amplified spontaneous emission and claim
that only the latter correspond to true lasing with coherent feedback.
Significance of these development consists in realization of the fact that weak
scattering by itself does not guarantee nonresonant feedback and that diffusion
model may not be applicable to active systems even when scattering is weak.
This point was reinforced in~\cite{vann01}, where it was shown that even  in a
weakly scattering active medium characterized by strong radiative leakage, as
well as spectral and spatial overlap of the modes, lasing from individual modes
can still occur. Clearly, this is only possible because even in a
weakly scattering system there exists a broad distribution of radiative
lifetimes emphasised by the onset of lasing.

Since observation of multipeak emission spectrum  in weakly scattering systems
often required enhanced pumping and its concentration within a rather small
volume of the sample, effects of the inhomogeneity of pumping on lasing
spectrum have also been studied in~\cite{wu06,wu07}. There it was concluded
that the inhomogeneity might play a role in promoting this phenomenon.

As a result of these developments the focus of theoretical research in the
field of random lasers has shifted from identifying special configurations
responsible for lasing to accepting that a random laser is a multimode system
and needs to be treated within the framework of a complete multimode lasing
theory. This theory can be developed along the lines of standard semiclassical
lasing theory, but it has to incorporate such features specific for random
lasers as a much larger role of radiative leakage of the modes and irregular
spatial dependence of the refractive index. While this theory is far from being
complete, a review of recent developments in this area appears to be useful. To
present such a summary is the main objective of this work, whose structure is
as follows. In section~\ref{modes} we discuss several alternative ways to
introduce modal description of strongly open systems. After presenting the
formal multimode theory in section~\ref{lastheor}, we turn to discussion of
recent experimental and numerical results in section~\ref{examp}.

\section{Modes of open systems
\label{modes}}

A problem of introducing modal description for open systems has a long history.
This problem is important not only for physics of lasers, but also as a first
necessary step toward quantizing electromagnetic field in open resonators. It
is not surprising, therefore, that there have been many alternative attempts to
introduce the system of modes suitable for separation time and coordinate
dependence of various physical quantities such as electric or magnetic fields.
The difficulty of the problem stems from the fact that openness makes the
problem non-Hermitian. Therefore, the standard recipes for introduction of
modes and quantization based on eigenvectors of Hermitian operators are not
applicable in this situation. In this section we review a few alternative
methods of defining electromagnetic modes of open systems suggested by various
authors.

In the Coulomb gauge $\nabla \cdot \l[\epsilon (\mbf r)\,  \mbf E
(\mbf r, t) \r] = 0$, electric field $\mbf E (\mbf r, t)$ is
governed by a wave equation
\eqn{
  \epsilon (\mbf r)\, \frac {\pd^2} {\pd t^2} \mbf E + \nabla \times
  \l(\nabla \times \mbf E \r) = 0,
\label{weqpas} }
where Gaussian units with the velocity of light in vacuum $c = 1$ are used.
Often, for the sake of simplicity, the vector nature of electromagnetic field
is neglected, which is possible if coupling between various polarizations in an
open spatially inhomogeneous system is insignificant. Then electric field is
described by the scalar wave equation
\eqn{
  \epsilon (\mbf r)\, \frac {\pd^2} {\pd t^2} E - \nabla^2 E = 0.
\label{weqscalpas} }
In this section we consider unloaded, or passive,
systems (without gain) and assume that all the decay of the field is
due to the openness, so that the dielectric constant
$\epsilon(\mbf r)$ is real. In order to demonstrate the variety of
approaches developed to deal with this problem we will discuss in
this section Fox-Li modes, subsection~\ref{sub:FoxLi}, quasimodes,
subsection~\ref{sub:nm}, system-and-bath approach,
subsection~\ref{sub:fp}, and the most recent development, constant-flux
modes, will be considered in subsection~\ref{sub:cf}.

\subsection{Transverse Fox-Li modes
\label{sub:FoxLi}}

Random lasers are, of course, not the first type of systems, in
which radiative losses cannot be neglected. Historically earliest,
the problem of introducing modal description in the presence of
strong radiative losses arose in connection with lasing properties
of so called unstable resonators~\cite{sieg89a}. Fox and
Li~\cite{fox68} suggested to define a mode of such a system as
a field distribution which reproduces itself after the wave
makes one complete round trip inside the resonator. One assumes that
there is a well-defined propagation direction of the wave inside the
resonator, so that a solution of \eq{weqscalpas} can be written in
the form
\eqn{
  E(\mbf r, t) = \mrm{Re}\, \l\{E (\mbf r) \exp \l[ \rmi \l(k z - \omega t \r)
  \r] \r\},
}
where $k = \omega \sqrt \epsilon$ [$\epsilon (\mbf r) =
\mrm{const}$] and $E (\mbf r)$ changes on a scale much longer than
$k^{-1}$. The distribution of the field in the transverse direction,
$E (\mbf r)$, is characterized by solutions of integral equation
\eqn{
  \int K(\mbf r_\perp, \mbf r'_\perp, z)\, \psi_n (\mbf r'_\perp, z)\, \rmd
  \mbf r'_\perp = \lambda_n \psi_n (\mbf r_\perp, z),
\label{FLint} }
which formally expresses the idea of the field
reproducibility and whose kernel is determined by optical
characteristics of the resonator. This equation has a form of an
eigenvector equation for a non-Hermitian linear  integral operator,
whose eigenvalue $|\lambda_n| < 1$ describes radiative losses after
a round trip. Since the respective eigenvectors are not orthogonal,
one has to introduce an adjoint operator describing propagation in the
backward direction as:
\eqn{
  \int K(\mbf r'_\perp, \mbf r_\perp, z)\, \phi_n (\mbf r'_\perp, z)\, \rmd
  \mbf r'_\perp = \lambda_n \phi_n (\mbf r_\perp, z).
\label{FLintadj}}
The two families of functions, called Fox-Li
modes, are biorthogonal, i.e.,
\eqn{
  \int \phi_m^* (\mbf r_\perp, z)\, \psi_n (\mbf r_\perp, z)\, \rmd \mbf
  r_\perp = \delta_{mn},
}
and can be used to construct a modal expansion of the field
\eqn{
  E (\mbf r) = \sum_n c_n (z)\, \psi_n (\mbf r_\perp, z).
}
If $\psi_n$ are normalized to unity, then
\eqn{
  K_n \equiv \int |\phi_n (\mbf r_\perp, z)|^2\, \rmd \mbf r_\perp \geq 1.
}
It was shown~\cite{sieg89a,sieg89b} that $K_n$ is the Petermann
excess-noise factor~\cite{pete79}, which describes the effect of the
openness of the resonator on the fundamental Schawlow-Townes
linewidth. While the Fox-Li modes played an important role in
understanding unstable resonators, their application to random lasers
and especially to the problem of field quantization is rather
limited.

\subsection{Quasimodes
\label{sub:nm}}

The idea of quasimodes was transferred to optics from quantum
physics. In order to describe scattering resonances in atomic and
molecular physics, it was proposed to solve Schr\"{o}dinger equation
with boundary conditions at infinity containing  only outgoing waves
and no incoming incident waves (so called Siegert or Gamow boundary
conditions)~\cite{mois98}. The solutions of the resulting
non-Hermitian eigenvector problem are characterized by complex
eigenvalues and eigenvectors diverging at infinity. The latter
circumstance makes it impossible to use these modes as a basis for
representation of the lasing field or to develop quantization
procedure representing field in the entire space. At the same time,
it was argued in~\cite{chin98} that under certain
realistic conditions it is possible to use these modes to represent
field inside the resonator. For both lasing and quantization problem
the field is needed everywhere, therefore there have been several
attempts to reformulate this problem in a way, which would produce a
meaningful system of basis vectors valid in the whole space.

One of the approaches, developed in~\cite{dutr00}, uses modes obeying
Siegert-Gamow boundary conditions to present the field inside the cavity, but
uses a different set of functions to describe outside field in order to avoid
the divergency problem. These authors argue that such modes describe
``natural'' evolution of the field inside the cavity after it was created and
allowed to evolve freely and, therefore, call them natural modes. The authors
employ an original definition of the biorthogonal inner product by presenting
left- and right-propagating components of these modes as components of a
two-dimensional spinor. This construction was studied in more details for a
one-dimensional cavity of constant refractive index open at one side and having
a perfect mirror at the other side. Its eigenfunctions satisfying the
outgoing-wave boundary conditions are of the form $\psi_n (x) \propto
\rme^{\rmi \kappa_n x} + r \rme^{- \rmi \kappa_n x}$, where $\kappa_n$ is
complex. Its adjoint function $\phi_n  (x) = [\psi_n (x)]^*$ matches an
incoming wave. The corresponding natural modes are then
\eqn{
  \Psi_n (x) \propto \l(\begin{array}{c} \rme^{\rmi \kappa_n x} \\
  r \rme^{- \rmi \kappa_n x} \end{array} \r), \quad
  \Phi_n (x) \propto \l(\begin{array}{c} r \rme^{\rmi \kappa^*_n x} \\
  \rme^{- \rmi \kappa^*_n x} \end{array} \r).
}
Natural modes for the external region are defined to have a real wavenumber~$k$,
but, if continued inside the cavity, they would not satisfy the Dirichlet
condition at the mirror. Both internal and external modes are biorthogonal to
their adjoints in the respective regions of space.

Since the natural modes form complete sets, they can be used to quantize the
field in the whole space. Namely, the amplitudes of the modes $\Psi_n (x)$ and
$\Phi_n (x)$ become cavity operators $a_n$ and $b_n$, respectively. Similar,
external operators $a(k)$ and $b(k)$ are defined. The system Hamiltonian
\eqn{
  H = H_{\mrm{in}} (\{a_n, b_n\}) + H_{\mrm{out}} (\{a(k), b(k)\})
\label{Hnat}
}
is a sum of internal and external contributions without cross-terms. Coupling
between internal and external waves in this formalism arises due to
noncommutativity of the internal and external operators. This circumstance
makes application of these modes not very convenient, therefore, it would be
interesting to try to introduce new commuting internal and external operators.
This would result in a Hamiltonian where coupling would enter explicitly in
the standard form of cross-terms.

\subsection{Feshbach projection in the  system-and-bath
quantization scheme
\label{sub:fp}}

Field quantization based on quasimodes~\cite{dutr00} (subsection~\ref{sub:nm})
is an example of the system-and-bath approach: one introduces separate
eigenmodes and field operators inside the open region designated as the
``system'' and in the surrounding free space (``bath''). This procedure can be
contrasted with the modes-of-the-universe approach, in which the modes are
defined in the whole space. The separation into a system and a bath often
provides a more clear physical description. For example, the complex energy of
a resonator state immediately yields its lifetime, while extracting this
information from the real continuous spectrum is less straightforward.

Feshbach projection technique~\cite{fesh62} offers a convenient way to perform
the system-and-bath quantization in a rather general setting of
equation~\eq{weqpas}~\cite{vivi03}, the only assumption being that $\epsilon
(\mbf r) = 1$ outside of a finite domain $\mc Q$ of arbitrary shape. The idea
of the method is to project the Hilbert space of the modes of the universe into
the Hilbert spaces of $\mc Q$ and its exterior $\mc P$ with appropriate
boundary conditions.

Separating variables in~\eq{weqpas} with an ansatz $\mbf E (\mbf r, t) =
\mrm{Re}\l[ \mbf E_{\omega} (\mbf r) \exp (- \rmi \omega t) \r]$ and
introducing a new, in general, vector-valued function
$\bphi_{\omega} (\mbf r) = \sqrt{\epsilon (\mbf
r)}\,  \mbf E_{\omega} (\mbf r)$ we can formulate an eigenvalue problem
\eqn{
  L \bphi_{\omega} \equiv \frac 1 {\sqrt{\epsilon (\mbf r)}} \nabla \times \l[
  \nabla \times \frac {\bphi_{\omega}} {\sqrt{\epsilon (\mbf r)}} \r] =
  \omega^2 \bphi_{\omega}
\label{Ldef}
}
for the Hermitian operator~$L$. After the projection into the subspaces $\mc Q$
and $\mc P$, an equivalent system of equations
\eqn{
  \l(\begin{array}{ll}
    L_{\mc{Q Q}} & L_{\mc{Q P}} \\
    L_{\mc{P Q}} & L_{\mc{P P}}
  \end{array}\r)
  \l(\begin{array}{l} \bmu_\omega \\ \bnu_\omega \end{array}\r)
  = \omega^2 \l(\begin{array}{l} \bmu_\omega \\ \bnu_\omega \end{array}\r)
}
for the restrictions $\bmu_\omega = \bphi_\omega\big|_{\mc Q}$ and $\bnu_\omega
= \bphi_\omega\big|_{\mc P}$  is obtained. The differential operators $L_{\mc{Q
Q}}$ and $L_{\mc{P P}}$ act in their domains and have the same bulk terms as
$L$~\eq{Ldef} and interface terms; $L_{\mc{Q P}}$~and $L_{\mc{P Q}}$ act at the
interface between $\mc Q$ and $\mc P$. There is a certain freedom in the
distribution of the interface terms between the operators, as well as in the
division of the whole space between $\mc Q$ and~$\mc P$. It is expected,
however, that physically relevant quantities are independent of this
choice~\cite{vivi04}. The eigenmodes of $L_{\mc{Q Q}}$ and $L_{\mc{P P}}$ are
determined from the equations
\eqn{
  L_{\mc{Q Q}}\, \bmu_\lambda = \omega_\lambda^2\, \bmu_\lambda, \quad
  L_{\mc{P P}}\, \bnu_m (\omega) = \omega^2\, \bnu_m (\omega),
\label{QPmodes}
}
where $\omega_\lambda$ is real and discrete, $\omega$~is real and continuous
and $m$ is a discrete channel index. The modes are required to obey the
boundary conditions that make the interface terms of the operators vanish. For
example, in one dimension these could be the Dirichlet/Neumann conditions on
the $\mc Q$/$\mc P$ side of the interface or vice versa. The eigenfunctions
$\bmu_\lambda (\mbf r)$ and $\bnu_m (\omega; \mbf r)$ of the Hermitian
operators form complete sets in their domains. Because of the special boundary
conditions, an expansion of a mode-of-the-universe $\bphi_\omega$ in terms of
these eigenmodes should deviate strongly from the exact result in a layer
around the interface. The width of the layer should go to zero as the number of
terms in the expansions becomes infinite.

The field is quantized in each domain separately by assigning the annihilation
operators $a_\lambda$ and $b_m (\omega)$ to the eigenmodes $\bmu_\lambda$
and~$\bnu_m (\omega)$. The system-and-bath Hamiltonian
\begin{eqnarray}
  \fl H = \sum_\lambda \hbar \omega_\lambda\, a_\lambda^\dag a_\lambda + \sum_m
  \int \rmd \omega \hbar \omega\, b_m^\dag (\omega)\, b_m (\omega) \nonumber
  \\
  + \hbar \sum_{\lambda m} \int \rmd \omega \l[W_{\lambda m} (\omega)\,
  a_\lambda^\dag b_m (\omega) + V_{\lambda m} (\omega)\, a_\lambda b_m (\omega)
  + \mrm{H.c.} \r],
\label{Hsb}
\end{eqnarray}
consists of the internal and external contributions and the interaction part
with $W_{\lambda m} (\omega) = \langle \bmu_\lambda |L_{\mc{Q P}}| \bnu_m
(\omega) \rangle$ and  $V_{\lambda m} (\omega) = \langle \bmu_\lambda^*
|L_{\mc{Q P}}| \bnu_m (\omega) \rangle$. In contrast to the
Hamiltonian~\eq{Hnat}, here the system and bath operators commute and the
interaction terms are explicit.  From the Heisenberg equations of motion for
$a_\lambda (t)$ and $b_m (\omega; t)$ one obtains the quantum Langevin equation
for $a_\lambda (\omega)$ in the frequency representation:
\eqn{
  \rmi \sum_{\lambda'} \l[\omega \delta_{\lambda \lambda'} - \Omega_{\lambda
  \lambda'} (\omega) \r] a_{\lambda'} (\omega) + F_\lambda (\omega) = 0,
\label{Laneq}
}
where $F_\lambda (\omega)$ is the bath noise operator. The non-Hermitian
frequency matrix
\begin{eqnarray}
  \Omega_{\lambda \lambda'} (\omega) = \omega_\lambda \delta_{\lambda \lambda'}
  - \rmi \pi \l(W W^\dag \r)_{\lambda \lambda'} (\omega) - \Delta_{\lambda
  \lambda'} (\omega),
\label{frmat} \\
  \l(W W^\dag \r)_{\lambda \lambda'} (\omega) \equiv
  \sum_m W_{\lambda m} (\omega)\, W_{\lambda' m}^* (\omega),
\end{eqnarray}
contains the imaginary damping term and the real frequency shift, often
disregarded. In the rotating-wave approximation, which is applicable when the
damping is much smaller than the typical frequencies, the $V_{\lambda m}
(\omega)$ contribution can be neglected.

Semiclassically, the electric field in the system (in $\mc Q$) is $\mbf
E_{\omega} (\mbf r) = \sum_\lambda a_\lambda (\omega) \bmu_\lambda (\mbf r)/
\sqrt{\epsilon (\mbf r)}$. Representing the matrix~\eq{frmat} via its
eigenvalues and biorthogonal left and right eigenvectors as $\Omega (\omega) =
\sum_k |r_k (\omega) \rangle \Omega_k (\omega) \langle l_k (\omega)|$ we can
construct quasimodes
\eqn{\fl
  \bpsi_k (\omega; \mbf r) = \langle \bmu (\mbf r) | r_k (\omega) \rangle,
  \quad \mbf E_{\omega} (\mbf r) = \sum_k \mbf E_k (\omega; \mbf r) =
  \sum_k a_k (\omega)\, \bpsi_k (\omega; \mbf r)/\sqrt{\epsilon (\mbf r)}
\label{normmod}
}
where $a_k (\omega) \equiv \langle l_k (\omega)|a (\omega)\rangle$ and $|\bmu
(\mbf r)\rangle$ [$|a (\omega)\rangle$] is a column of $\bmu_\lambda (\mbf r)$
[$a_\lambda (\omega)$]. These modes depend on the real frequency $\omega$
imposed by the exterior. With the help of \eq{Laneq} and the bath correlation
function at zero temperature $\langle F_\lambda (\omega)\, F_{\lambda'}^\dag
(\omega') \rangle = \l(W W^\dag \r)_{\lambda \lambda'}\! (\omega)\, \delta
(\omega - \omega')$~\cite{hack03} one obtains the (bath averaged) field
correlation in the mode~$k$~\cite{zait_up}:
\eqn{\fl
  \langle \mbf E_k (\omega; \mbf r) \cdot \mbf E_k^* (\omega'; \mbf r) \rangle
  = \frac {\langle l_k (\omega)| W W^\dag (\omega) |l_k (\omega) \rangle}
  {\l[\omega - \omega_k (\omega) \r]^2 + \kappa_k^2 (\omega)}
  \frac {|\bpsi_k (\omega; \mbf r)|^2} {\epsilon (\mbf r)}
  \delta (\omega - \omega'),
}
where $\omega_k (\omega) \equiv \mrm{Re}\, \Omega_k (\omega)$ and $\kappa_k
(\omega) \equiv \mrm{Im}\, \Omega_k (\omega)$. Thus, the modes $\mbf E_k
(\omega; \mbf r)$ indeed describe a leaking field with the frequency given by
the equation $\omega = \omega_k (\omega)$ and the decay rate~$\kappa_k
(\omega)$, as is expected from quasimodes.

\subsection{Constant-flux states
\label{sub:cf}}

So called constant-flux (CF) states were introduced in~\cite{ture06}
as an attempt to construct a system of basis
functions suitable for modal expansion of the field inside an open
resonator, which would also allow easily calculate field outside of
the resonator. While these modes are, indeed, very convenient for
semiclassical lasing theory, it is not clear at the present time if
they can be used for quantization of the field. The CF modes are
very similar to the modes~\cite{vivi03} of the previous
subsection~\ref{sub:fp}: in the interior of the cavity region
$\mc Q$ they satisfy the eigenvalue equation
\eqn{
  \frac 1 {\sqrt{\epsilon (\mbf r)}} \nabla \times \l[ \nabla \times \frac
  {\wtilde \bpsi_k (\omega)} {\sqrt{\epsilon (\mbf r)}} \r] = \wtilde
  \Omega_k^2 (\omega)\, \wtilde \bpsi_k (\omega),
\label{eq:cfin}
}
where  $\wtilde \Omega_k$ is an eigenfrequency and in
the exterior domain~$\mc P$ the equation for these modes takes the
form
 \eqn{
  \nabla \times \l[ \nabla \times \wtilde \bpsi_k (\omega) \r] = \omega^2\,
  \wtilde \bpsi_k (\omega)
\label{eq:cfout}
}
where $\omega$ is a real external spectral parameter, which does not coincide
with the eigenfrequency~$\wtilde \Omega_k$. However, equations~\eq{eq:cfin}
and~\eq{eq:cfout} are complemented by outgoing-wave boundary conditions at
infinity and continuity conditions at the boundary of the resonator. One can
think of $\omega$ as of frequency of some external force exciting the modes of
the cavity, or alternatively, as a spectral parameter of temporal Fourier
transform used to convert the problem from time to frequency domain. This
spectral parameter becomes an integration variable when inverse transform back
to time domain is carried out. Description of the outside modes by
equation~\eq{eq:cfout} not only allows to introduce nondiverging energy flux,
but also ensures the biorthogonality of the internal modes. Indeed,
electromagnetic boundary conditions are of mixed type and contain dependence on
the eigenfrequencies. As a result the orthogonality between adjoint modes is
destroyed, and transition to \eq{eq:cfout} is a way to restore it. Similar
approach was described in the case of elastic waves in~\cite{mors53}. It was
shown in~\cite{ture06} that modes defined this way form a complete set in the
domain~$\mc Q$, which can be complemented by a biorthogonal set of adjoint
modes. The advantage of these states compared to quasimodes obeying
Siegert-Gamow boundary conditions is that the former remain finite at infinity
and describe constant flux of energy coming out from the resonator (hence,
``constant flux'' states). There is no need to introduce special outside modes,
the exterior field is calculated by simply matching outgoing field with the
cavity field. This is particularly convenient for calculation of light emission
from open resonators.

The CF wavefunctions and their biorthogonal adjoint functions~$\wtilde \bphi_k
(\omega; \mbf r)$ provide spectral representation of the interior
Green-function operator~\cite{ture06} satisfying outgoing boundary conditions
at infinity, which can be written in the form
\eqn{
  [G_{\mc{QQ}} (\omega; \mbf r, \mbf r')]_{\alpha \alpha'} = \sum_k \frac
  {[\wtilde \bpsi_k (\omega; \mbf r) ]_\alpha\, [\wtilde \bphi_k^*
  (\omega; \mbf r')]_{\alpha'}} {\omega^2 - \wtilde \Omega_k^2 (\omega)}.
\label{GF}
}
The indices $\alpha, \alpha' = x,y,z$ label the polarization of vector-valued
field. The Green function satisfying the same boundary conditions can be
defined in the system-and-bath-approach as~\cite{vivi03} $G_{\mc{QQ}} (\omega)
= [\omega^2 - L_{\mrm{eff}} (\omega)]^{-1}$, where the differential operator
$L_{\mrm{eff}}$ has the form
\eqn{
  L_{\mrm{eff}} (\omega) = L_{\mc{Q Q}} + L_{\mc{Q P}}\, (\omega^2 - L_{\mc{P
  P}} + \rmi \varepsilon)^{-1} L_{\mc{P Q}}.
}
Thus, $\wtilde \bpsi_k (\omega)$ are eigenfunctions of $L_{\mrm{eff}} (\omega)$
with the eigenvalues~$\wtilde \Omega_k^2 (\omega)$. In fact, it was shown
in~\cite{vivi04} for the one-dimensional case, that the interface terms of
$L_{\mrm{eff}} (\omega)$ disappear precisely when the function it acts upon
satisfies the outgoing boundary conditions. The poles of the Green function,
which correspond to scattering resonances, are found by analytical continuation
of $\omega$ into the complex plane such that the equation $\omega^2 =
\wtilde\Omega_k^2 (\omega)$ is satisfied. At the same time, in the case of CF
states, $\omega$~is always real and, if it is fixed by some external conditions
(for instance, it can be frequency of incident radiation tuned to be in
resonance with the cavity mode), the respective resonant eigenfrequency of the
CF state obeys a different equation: $\mrm{Re}\, [\Omega_k(\omega)]=\omega$.

It is interesting to compare CF modes with those obtained in the
system-and-bath approach described in subsection~\ref{sub:fp}. The
first step is to write $L_{\mrm{eff}}$ as a matrix in the eigenbasis of
$L_{\mc{Q Q}}$ [see~\eq{QPmodes}]:
\eqn{
  L_{\mrm{eff}} (\omega) \simeq \Omega_0^2 - 2 \sqrt{\Omega_0} \l[\rmi \pi W
  W^\dag (\omega) + \Delta (\omega) \r] \sqrt{\Omega_0},
}
where $\Omega_0$ is the real diagonal matrix of~$\omega_\lambda$. The
difference between the matrix $\Omega^2 (\omega)$~\eq{frmat} and this
matrix reads
\eqn{\fl
  \Omega^2 (\omega) - L_{\mrm{eff}} (\omega) = [\Delta \Omega (\omega),
  \Omega_0] + \Delta \Omega^2 (\omega), \quad \Delta \Omega (\omega) \equiv
  \Omega (\omega) - \Omega_0,
}
where the diagonality of $\Omega_0$ was used. This difference is of the second
order in the small parameters $|\Delta \Omega_{\lambda
\lambda'}/|\omega_\lambda|$ and $|\omega_\lambda -
\omega_{\lambda'}|/|\omega_\lambda|$. Therefore, the CF eigenfrequencies
and eigenfunctions are
expected to become close to $\Omega_k (\omega)$ and $\bpsi_k (\omega)$ as the
above ratios decrease.

\subsection{Open resonators in random-matrix theory
\label{RMT}}

Explicit construction of quasimodes along the lines of subsections
\ref{sub:nm}-\ref{sub:cf} in irregular systems may require a fair amount of
computations. At the same time, one is often interested in statistical
description of ensembles of such systems. It is well known that certain
statistical characteristics of chaotic cavities and diffusive media in the
short-wavelength limit are universal. This means that these properties depend
only on the symmetries and, possibly, the boundary conditions, but not on the
details of the spatial distribution of the refractive index. The universality
justifies modelling eigenfrequencies and eigenfunctions of these systems using
appropriate statistical ensembles of random matrices, which can be easier to
handle analytically or numerically than physical systems. A review of
random-matrix theory (RMT) in open cavities is given in~\cite{fyod97}.

A convenient starting point is the following, rather general representation of
the scattering matrix of an open system:
\eqn{
  S (\omega) = \1 - 2 \rmi \pi W^\dag (\omega - \Omega_{\mrm{eff}})^{-1}\, W,
  \quad \Omega_{\mrm{eff}} = \Omega_0 - \rmi \pi W W^\dag.
\label{scatmat}
}
$S (\omega)$ is an $M \times M$ matrix in the channel space.
It transforms a column of incoming amplitudes into a column of outgoing
amplitudes in some basis of the channel states. $\Omega_{\mrm{eff}}$~is called
effective Hamiltonian in electronic systems. It is analogous to the frequency
matrix~\eq{frmat}. $\Omega_0$~is an $N \times N$ Hermitian frequency matrix of
the closed resonator; its eigenvalues are the eigenfrequencies of the cavity.
The $N \times M$ matrix $W$ describes coupling between the resonator and the
open channels. In the limit usually considered in RMT, $N \to \infty$, the
frequency dependence of $W$ is neglected.

RMT prescribes that, in order to model a generic chaotic system with
time-reversal symmetry, the real symmetric matrix $\Omega_0$ should be taken
from the Gaussian orthogonal ensemble. Without loss of generality, the diagonal
(off-diagonal) elements of $\Omega_0$ are drawn from a normal distribution with
zero mean and the variance of $2/N$~($1/N$). In the limit $N \to \infty$ the
eigenvalues are distributed according to the Wigner semicircle law
\eqn{
  g (\omega) = \frac 1 \pi \sqrt{1 - \frac {\omega^2} 4}, \quad -2 \leq \omega
  \leq 2,
}
where $g (\omega)$ is normalized to unity. In the absence of direct coupling
between the channels, matrix $S (\omega)$ should become diagonal after the
ensemble average. This can be achieved simply by taking fixed (for all members
of the ensemble) elements $W_{nn} \equiv \sqrt{\gamma_n / \pi} > 0$ for $n \leq
M$ and $W_{nm} = 0$ otherwise (the limit $M \ll N$ is assumed). Thus, the
interaction matrix $\pi W W^\dag$ is diagonal with $M$ nonzero
elements~$\gamma_n$. The coupling strength between the resonator and the
continuum is characterized by the transmission coefficients
\eqn{
  T_n = 1 - |\langle S_{nn} (\omega) \rangle|^2 = 2 \l[1 + \frac {\gamma_n +
  \gamma_n^{-1}} {2 \pi g (\omega)} \r]^{-1}.
}
In particular, the coupling is the strongest for $\gamma_n = 1$. Since the
coupling depends on the density~$g (\omega)$, one has to be careful not to mix
the statistics from different regions of~$\omega$ or introduce appropriate
scaling in order to obtain universal results.

Quasimodes of an open chaotic resonator can be modelled via the resonances
of $S (\omega)$, which are the eigenmodes of~$\Omega_{\mrm{eff}}$. Statistical
properties of the resonant wavefunctions will follow from the surmise that
values of an eigenfunction of a closed chaotic cavity are uncorrelated Gaussian
random variables at points spaced more than a wavelength apart~\cite{berr77}.

Several papers extend the calculations~\cite{sieg89a,sieg89b} of the Petermann
factor~\cite{pete79} to chaotic systems, where, in particular, there is no
separation into longitudinal and transverse modes. In the case of chaotic
cavity with an opening~\cite{patr00,frah00}, the RMT effective Hamiltonian
$\Omega_{\mrm{eff}}$~\eq{scatmat} is modified to account for the amplifying
medium in the resonator by adding a correction $\rmi \gamma_{\mrm a}/2$ to its
eigenvalues, where $\gamma_{\mrm a} > 0$ is the amplification rate. As
$\gamma_{\mrm a}$ is increased from zero, the eigenvalues shift upwards in the
complex plane (the gain is assumed to be the same for all relevant
frequencies). The first eigenvalue touching the real axis defines the lasing
threshold. The Petermann factor is shown to be
\eqn{
  K = \langle l | l \rangle \langle r | r \rangle
\label{K}
}
where $| l \rangle$ ($| r \rangle$) is the left (right)
eigenvector of $\Omega_{\mrm{eff}}$ (for $\gamma_{\mrm a} = 0$). A
supersymmetric calculation~\cite{frah00} reveals that $\langle K \rangle$
scales as the square root of the number of channels in the opening. If the
system has a time-reversal symmetry, the matrix $\Omega_{\mrm{eff}}$ is
symmetric, i.e., $W$ is real. Then one can impose the condition $| l \rangle =
| r \rangle^*$, which leads to an alternative expression
\eqn{
  K_{\mrm{TRS}} =  \langle r | r \rangle^2.
\label{KTRS}
}

In a chaotic dielectric resonator (a domain of uniform refractive index $n >
1$ surrounded by a medium with $n = 1$) light can leak anywhere along the
boundary. Hence, the number of output channels~$M$ scales with the size~$N$ of
$\Omega_{\mrm{eff}}$, breaking the standard RMT assumption $M \ll N$. A
combination of RMT with the Fresnel laws~\cite{keat08} produces a scattering
matrix
\eqn{
  S (\omega) = - R + T F (\omega) \l[\1 - R F (\omega) \r]^{-1} T,
}
where $F (\omega)$ is the intracavity propagator and $R$ and $T$ are
diagonal matrices of reflection and transmission. The Petermann factor is of
the form~\eq{K}, but the eigenvectors refer to the matrix~$R F (\omega)$.
After a symmetrization procedure, the form~\eq{KTRS} can also be obtained if
time-reversal symmetry is present. Comparison of the RMT calculations with the
quantum-kicked-rotator model points at loss of universality as $n$ becomes
close to unity, i.e., in a strongly open resonator.

It should be noted that the near-threshold
treatment~\cite{patr00,frah00,keat08} yields the value of Petermann factor
which is twice as large as the correct result. This happens because phase
diffusion (which mainly contributes to the line width) is amended at the
threshold by the amplitude fluctuations. The correct prefactor was inserted
``by hand'' in equations \eq{K} and~\eq{KTRS}. A~linearization of the quantum
Langevin equations far above the threshold~\cite{hack02} yields again the
result~\eq{K}, this time with the proper prefactor.

\section{Multimode laser theories for open and irregular systems
\label{lastheor}}

As it was mentioned in introductory  section~\ref{sec:intro}, the current trend
in theoretical description of random lasing consists in extending standard
semiclassical mutlimode lasing theory~\cite{sarg74,hake84} to situations
specific for random lasing: modes with broad distribution of radiative
lifetimes and irregular spatial patterns, inhomogeneity of the background
refractive index, large number of lasing modes emitting in the regime when
pumping significantly exceeds the threshold value. While it is true that the
multimode lasing was considered one of the signatures of coherent feedback, the
discussion of this issue in earlier works related to strongly scattering
systems was mostly concerned with a rather trivial situation, when multiple
modes originated from different non-overlapping cavities (localized states),
which were assumed to be in the single-mode regime due to mode
competition~\cite{jian00,souk02,cao03b}. The only nontrivial multimode effect
reported in strongly scattering systems was observation of mode coupling
in~\cite{cao03b}, however, an attempt of its explanation using a two-mode
lasing theory given in~\cite{jian04} relied on incorrect form of rate equations
(this point will be explained later in this section).

One of the first attempts to apply  multimode lasing theory to chaotic lasers
was undertaken in~\cite{misi98}, where equations of standard semiclassical
laser theory were combined with ideas of random-matrix theory concerning
statistical properties of eigenmodes of chaotic resonators. Cavities studied
in~\cite{misi98} were characterized by a relatively long radiative lifetimes of
the modes. Statistical properties of random lasers with strong radiative losses
were studied in~\cite{hack05,zait06,zait07} using combination of random-matrix
theory with the method of Feshbach projection. A statistical study of
one-dimensional strongly open random lasers based on calculations of lasing
modes ``from the first principles'' rather than on phenomenological
random-matrix-type considerations was carried out in~\cite{zait09}.

The role of nonuniformity  of the refractive index in formation of lasing modes
and its consequences for lasing dynamics was discussed
in~\cite{deyc05a,deyc05b}, where the idea that lasing modes can be
significantly different from modes of cold cavities and must be determined
selfconsistently was formulated. Application of the
theory~\cite{deyc05a,deyc05b} to random lasers was, however, limited because of
restriction of the selfconsistency requirements to the linear regime only and
neglect of radiative losses of the modes. Both these limitations were removed
in~\cite{ture06,ture07,ture08} (see also recent review article~\cite{ture09}),
where spatial structure of lasing modes  and lasing frequencies were determined
selfconsistently from fully nonlinear theory. The approach developed in these
papers is based on three main ingredients: (i) use of CF states to incorporate
radiative losses of the system, (ii) neglecting population pulsation, which
allowed to take into account nonlinear interactions up to infinite order in
field intensity and (iii) determination of lasing modes and their frequencies
selfconsistently. Results of those works revealed the presence of strong
effects related to nonlinear interaction between modes in systems with strongly
overlapping (both spectrally and spatially) modes. However, since this method
is based on numerical computation of fixed points of a certain nonlinear map,
which, as it often happens in nonlinear systems, might have multiple stable and
metastable points, the usefulness of this approach might be limited to systems
with relatively large intermode spacing. Its applicability to truly diffusive
random lasers is difficult to assess from the data published
in~\cite{ture06,ture07,ture08,ture09}, since they do not contain any
information about mean free path, $l_{\mrm{mfp}}$, of light in studied
structures. Diffusive regime arises only when relations $R \gg l_{\mrm{mfp}}
\gg \lambda$, where $\lambda$ is the wavelength of light and $R$ is the size of
the sample, are satisfied, which requires much larger samples than the
condition $R \gg \lambda$, fulfilled in samples studied
in~\cite{ture06,ture07,ture08,ture09}.

In this and the following section we derive a semiclassical multimode lasing
theory taking into account recent achievements discussed above and present some
examples of its application to situations relevant for random lasers.

\subsection{Semiclassical laser equations}

A starting point for the semiclassical description of
lasers~\cite{sarg74,hake84} is the wave equation for the electric
field~\eq{weqpas} with the polarization $\mbf P (\mbf r, t)$ as a source that
generates the field ($c = 1$):
\eqn{
  \epsilon (\mbf r)\, \frac {\pd^2} {\pd t^2} \mbf E + \nabla \times \l(\nabla
  \times \mbf E \r) = - 4 \pi \frac {\pd^2} {\pd t^2} \mbf P (\mbf r, t).
\label{weqact}
}
In the simplest model, the polarization is produced by two-level active atoms
and obeys the equation
\eqn{
  \l(\frac {\pd^2} {\pd t^2} + 2 \gamma_\perp \frac \pd {\pd t} + \nu^2 \r)
  \mbf P =  -2 \nu \frac {d^2} \hbar \mbf E (\mbf r, t)\, \Delta n (\mbf r, t),
\label{poleq}
}
where $\Delta n (\mbf r, t)$ is the population-inversion density, $d$~is the
magnitude of the atomic dipole matrix element, $\nu$~is the atomic transition
frequency (homogeneous broadening is assumed) and $\gamma_\perp$~is the
polarization decay rate. The population inversion, in turn, depends on the
electric field and the polarization,
\eqn{
  \frac \pd {\pd t} \Delta n - \gamma_\parallel [\Delta n_0 (\mbf r, t) -
  \Delta n] = \frac 2 {\hbar \nu} \mbf E (\mbf r, t) \cdot \frac \pd {\pd t}
  \mbf P (\mbf r, t).
\label{inveq}
}
If the right-hand side vanishes, $\Delta n$~relaxes with the rate
$\gamma_\parallel$ to the unsaturated population invertion~$\Delta n_0 (\mbf r,
t)$, which is determined by the pump.

The coupled equations \eq{weqact}-\eq{inveq} yield, in principle,
distribution of the electric field in the system, if~$\Delta n_0
(\mbf r, t)$ is given. Unlike the traditional derivation of lasing
equations, which is usually done in time domain, we find it more convenient to
proceed using frequency representation.  We introduce the Fourier
transforms
\begin{eqnarray}
  \mbf E (\mbf r, t)& =& \frac 1 \pi \mrm{Re} \int_0^\infty \rmd \omega\, \mbf
  E_\omega (\mbf r)\,  \rme^{-\rmi \omega t}, \\
  \mbf P (\mbf r, t) &=& \frac 1 \pi \mrm{Re} \int_0^\infty \rmd \omega\, \mbf
  P_\omega (\mbf r)\,  \rme^{-\rmi \omega t}, \\
  \Delta n (\mbf r, t)& =& \frac 1 {2 \pi} \int_{-\infty}^\infty \rmd \omega
  \Delta n_\omega (\mbf r)\, \rme^{-\rmi \omega t}.
\end{eqnarray}
in the form chosen to facilitate application of the rotating-wave
approximation. In addition, we assume that the time dependence of the field and
polarization is determined by fast oscillations with frequencies which are
close to atomic frequency~$\nu$ and residual slow time dependence. In the
frequency domain this means that only Fourier components $\mbf E_\omega$ and
$\mbf P_\omega$ with $\omega$ in the small vicinity of $\nu$
contribute significantly to the
dynamics. Therefore, after performing Fourier transform of
\eq{weqact}-\eq{inveq}, one can neglect terms of the order of $(\omega-\nu)^2$
in $\omega^2=(\omega-\nu+\nu)^2 \approx \nu^2 + 2 \nu (\omega - \nu)$ and write
down the resulting equations as
\begin{eqnarray}
  \fl
  - \epsilon (\mbf r)\, (-\nu^2 + 2 \nu \omega)\, \mbf E_\omega + \nabla \times
  (\nabla \times \mbf E_\omega) = 4 \pi \nu^2 \mbf P_\omega,
\label{weqactft}\\
  \fl
  [-\rmi (\omega - \nu) + \gamma_\perp ] \mbf P_\omega = - \rmi \frac
  {d^2} {2 \pi \hbar} \int_0^\infty \rmd \omega'\, \mbf E_{\omega'} \Delta
  n_{\omega - \omega'},
\label{poleqft}\\
  \fl
  (-\rmi \omega + \gamma_\parallel) \Delta n_\omega = 2 \pi \gamma_\parallel
  \Delta n_0 (\mbf r)\, \delta (\omega) - \frac \rmi {\pi \hbar} \int_0^\infty
  \rmd \omega' (\mbf E_{\omega' - \omega}^* \cdot \mbf P_{\omega'} - \mbf
  E_{\omega' + \omega} \cdot \mbf P_{\omega'}^*),
\label{inveqft}
\end{eqnarray}
where we  assumed that the pump $\Delta n_0 (\mbf r)$ is time independent.

In the linear approximation~\cite{deyc05a,deyc05b} (valid below and not far
above the lasing threshold) we neglect the quadratic terms in~\eq{inveqft} and
use
\eqn{
  \Delta n_\omega^{(0)} = 2 \pi \Delta n_0 (\mbf r)\,
  \delta (\omega)
}
in~\eq{poleqft}. Polarization in this approximation is given by
\eqn{
  \mbf P_\omega^{(1)} = - \rmi
  (d^2 / \hbar \gamma_\perp) D(\omega) \Delta n_0 (\mbf r)\, \mbf
  E_\omega
}
and, when substituted to the right-hand side of~\eq{weqactft}, yields the
following equation for the electric field:
\eqn{
  \fl
  - \epsilon (\mbf r)\, (-\nu^2 + 2 \nu \omega)\, \mbf E_\omega + \nabla \times
  (\nabla \times \mbf E_\omega) = -4 \pi \rmi \frac {d^2 \nu^2} {\hbar
  \gamma_\perp} D (\omega)\, \Delta n_0 (\mbf r)\, \mbf E_\omega,
\label{linlas}
}
where $D (\omega) \equiv [1 - \rmi (\omega - \nu)/\gamma_\perp]^{-1}$.

One can use any of the system of modes $\bpsi_k (\omega; \mbf r)$ discussed in
subsections \ref{sub:nm}-\ref{sub:cf} and their adjoint modes $\bphi_k (\omega;
\mbf r)$ with respective eigenfrequencies $\Omega_k (\omega)$ (we use the same
notation for all types of modes) in order to generate modal
expansion~\eq{normmod} of electric field in equation~\eq{linlas}. The expansion
coefficients are found using the biorthogonal functions~as
\eqn{
  a_k (\omega) = \int \rmd \mbf r\, \sqrt{\epsilon (\mbf r)}\, \bphi_k^*
  (\omega; \mbf r) \cdot \mbf E_\omega (\mbf r).
\label{ak}
}
Then \eq{linlas} is reduced to the matrix eigenvalue problem
\eqn{
  \sum_{k'}\, [\omega \delta_{kk'} - \wbar \Omega_{kk'} (\omega)]\, a_{k'}
  (\omega) = 0
\label{linmod}
}
with
\begin{eqnarray}
  \wbar \Omega_{kk'} (\omega) = \Omega_k (\omega) \delta_{kk'} + \rmi 2 \pi \nu
  \frac {d^2} {\hbar \gamma_\perp} D (\omega) V_{kk'} (\omega),
\label{Ombar}\\
  V_{kk'} (\omega) = \int \rmd \mbf r\, \bphi_k^* (\omega; \mbf r) \cdot
  \bpsi_{k'} (\omega; \mbf r) \frac {\Delta n_0 (\mbf r)} {\epsilon (\mbf r)}.
\end{eqnarray}
The matrix $V_{kk'} (\omega)$ becomes diagonal for uniform $\Delta n_0$
and~$\epsilon$. The lasing thresholds and frequencies at the thresholds are
determined from the system of equations $\omega = \mrm{Re}\, \wbar \Omega_k
(\omega)$ and $\mrm{Im}\, \wbar \Omega_k (\omega) = 0$, where $\wbar \Omega_k
(\omega)$ are eigenvalues of~$\wbar \Omega_{kk'} (\omega)$.

\subsection{Third-order theory}

Nonlinear effects can be included by iterating equations \eq{poleqft}
and~\eq{inveqft}, with the field as a small parameter~\cite{sarg74,hake84}.
Namely, $\mbf P_\omega^{(1)}$ is inserted in~\eq{inveqft} to obtain the
correction~$\Delta n_\omega^{(2)}$, which, in turn, is used in~\eq{poleqft} to
yield the contribution to the polarization~$\mbf P_\omega^{(3)}$, of the third
order in the electric field.

Assuming that lasing modes exist, we can present field as a sum of oscillating
terms with slowly varying amplitudes,
\eqn{
  \mbf E (\mbf r, t) = \mrm{Re} \l[\sum_l \mbf E_l (\mbf r, t)\, \rme^{- \rmi
  \omega_l t} \r], \quad \mbf E_\omega (\mbf r) = \sum_l \mbf E_l (\omega -
  \omega_l; \mbf r),
\label{lasmod}
}
and positive frequencies~$\omega_l$ close to~$\nu$, which are to be determined
selfconsistently. If the amplitudes $\mbf E_l (\mbf r, t)$ vary slowly on the
scale of~$\omega_l$, their Fourier transforms $\mbf E_l (\omega - \omega_l;
\mbf r)$ are strongly peaked at~$\omega_l$. The slow-varying-amplitude
approximation in the frequency domain amounts to the replacement $\mbf E_l
(\omega - \omega_l; \mbf r) \to \pi \mbf E_l(\mbf r, t)\, \delta(\omega -
\omega_l)$ in nonlinear terms before performing frequency integrals when
transforming to the time representation. Further, we average out interference
terms, oscillating at the beat frequencies, and neglect mode degeneracies
($|\omega_l - \omega_m| \gg |\dot {\mbf E}_l| / |\mbf E_l|$). After the
nonlinear correction~$\mbf P_\omega^{(3)}$ is added to the right-hand side
of~\eq{weqactft} and the linear contribution is diagonalized according
to~\eq{linmod}, we arrive at the third-order lasing equations
\begin{eqnarray}
  \fl
  \l\{\frac \rmd {\rmd t} + \rmi \l[\wbar \Omega_k (\omega_m) - \omega_m \r]
  \r\} \wbar a_{km} (t) = \nonumber \\
  \fl
  - \frac {\pi \nu} {\hbar \gamma_\parallel} \l( \frac {d^2} {\hbar
  \gamma_\perp} \r)^2 D(\omega_m)  \int \rmd \mbf r\, \frac {\Delta n_0 (\mbf
  r)} {\sqrt{\epsilon (\mbf r)}}\, \wbar {\bphi}_k^* (\omega_m; \mbf r) \cdot
  \sum_{l = 1}^{N_{\mrm m}} \l\{ 2 \mrm{Re}\, [D(\omega_l)]\, \mbf E_m (\mbf
  r, t)\, |\mbf E_l (\mbf r, t)|^2 \r. \nonumber \\
  \fl
   \l.+\, (1 - \delta_{ml})\, D_\parallel (\omega_m - \omega_l)\, [D(\omega_m)
  + D^*(\omega_l)]\, \mbf E_l (\mbf r, t)\, [\mbf E_l^* (\mbf r, t) \cdot \mbf
  E_m (\mbf r, t)] \r\}.
\label{3rdord}
\end{eqnarray}
$a_{km}$ is the amplitude of the $k$-th component of the $m$-th lasing mode in
the basis of eigenfunctions of linearized problem~\eq{linlas} and~\eq{linmod}:
\eqn{
  \mbf E_m (\mbf r, t) = \sum_{k = 1}^{N_{\mrm b}} \wbar a_{km} (t)\, \wbar
  {\bpsi}_k (\omega_m; \mbf r) / \sqrt{\epsilon (\mbf r)}$, \quad $m = 1,
  \ldots, N_{\mrm m}.
}
This is a system of $N_{\mrm b} \times N_{\mrm m}$ equations, where $N_{\mrm
b}$ is the basis size and $N_{\mrm m}$ is the number of lasing modes. The terms
proportional to $D_\parallel (\omega) \equiv (1 - \rmi \omega /
\gamma_\parallel)^{-1}$ arise from the population pulsations at the beat
frequency $\omega_m - \omega_l$ interfering in~\eq{poleqft} with the
oscillations at the frequency~$\omega_l$ (and, thus, surviving the averaging).
In the stationary regime ($\rmd \bar{a}_{km}/\rmd t=0$), the amplitudes~$\wbar
a_{km}$ and frequencies~$\omega_m$ can be determined from equations~\eq{3rdord}
by an iteration procedure (see section~\ref{aot}).

The number of equations in system \eq{3rdord} can be reduced to~$N_{\mrm b}$ if
one assumes that the wavefunctions of the lasing modes are given by the linear
approximation~\eq{linlas} and do not have to be determined selfconsistently
from the nonlinear equations~\eq{3rdord}. This is the approximation used
in~\cite{deyc05a,deyc05b}, where it was justified by the fact that such a
partially (in the linear approximation) selfconsistent treatment still allows
one to eliminate fast oscillating terms in the nonlinear polarization and to
introduce the slow-varying-amplitude approximation resulting in rate equations
for the respective amplitudes. In this linearly selfconsistent approximation
the amplitudes can be presented as $\wbar a_{km} (t) = \wbar a_m (t)\,
\delta_{km}$ and the lasing equations for a scalar field take the form
(cf.~\cite{hack05,zait07,zait09})
\begin{eqnarray}
  \fl
  \l\{\frac \rmd {\rmd t} + \rmi \l[\wbar \Omega_m (\omega_m) - \omega_m \r]
  \r\} \wbar a_m = - \frac {\pi \nu} {V \hbar \gamma_\parallel} \l( \frac {d^2}
  {\hbar \gamma_\perp} \r)^2 D(\omega_m)\, \wbar a_m \sum_l B_{ml}\, |\wbar
  a_l|^2 \nonumber \\
  \fl
  \times \l\{ 2  \mrm{Re}\, [D(\omega_l)] +\, (1 - \delta_{ml})\, D_\parallel
  (\omega_m - \omega_l)\, [D(\omega_m) + D^*(\omega_l)] \r\},
\label{3rdordsimpl}\\
  \fl
  B_{ml} = V \int \rmd \mbf r\, \frac {\Delta n_0 (\mbf r)} {[\epsilon (\mbf
  r)]^2}\, \wbar {\phi}_m^* (\omega_m; \mbf r)\, \wbar {\psi}_m (\omega_m;
  \mbf r)\, \l|\wbar {\psi}_l (\omega_l; \mbf r) \r|^2,
\label{corr}
\end{eqnarray}
where $V$ is the volume. Equations~\eq{3rdordsimpl} generalize the standard
third-order semiclassical theory with the saturation (hole-burning)
terms~\cite{sarg74,hake84} to the case of strongly open and irregular systems.
Separating real and imaginary parts of this equation one obtains rate equations
for intensities $I_m = |\wbar a_m|^2$ of the modes and an equation for lasing
frequencies:
\begin{eqnarray}
  \fl
  \l\{\frac \rmd {\rmd t} - 2\, \mrm{Im} \l[\wbar \Omega_m (\omega_m) \r] \r\}
  I_m = - \frac {2 \pi \nu} {V \hbar \gamma_\parallel} \l( \frac {d^2} {\hbar
  \gamma_\perp} \r)^2 I_m\, \mrm{Re} \l[D(\omega_m) \sum_l\, (\cdots) \r],
\label{rateeq}\\
  \fl
  \mrm{Re} \l[\wbar \Omega_m (\omega_m) \r]  - \omega_m = - \frac {\pi \nu} {V
  \hbar \gamma_\parallel} \l( \frac {d^2} {\hbar \gamma_\perp} \r)^2 \mrm{Im}
  \l[D(\omega_m) \sum_l\, (\cdots) \r],
\end{eqnarray}
where the sum $\sum_l\, (\cdots)$ appearing in~\eq{3rdordsimpl} depends on
intensities of all lasing modes. These rate equations do not contain any linear
coupling terms contrary to the assumption made in~\cite{jian04}.

These equations show that all statistical characteristics of laser emission
(frequency, threshold and intensity distributions) are determined by certain
integrals involving eigenfunctions of cold cavities. Transition from strong to
weak scattering manifests itself in changing statistical characteristics of
respective quantities. However, in spite of large amount of work on
wavefunction statistics in closed systems, the statistical properties of self-
and cross-saturation coefficients in open resonators have not yet been studied.
At the same time, it is clear now that this statistics is responsible for
various regimes of behavior of random lasers~\cite{zait09}. We will discuss
this point in more details in subsection~\ref{modstat}.

\subsection{All-order nonlinear theory in the time-independent
population approximation
\label{aot}}

It is possible to obtain lasing equations valid in all orders in the electric
field in a closed form if one neglects time dependence of  the population
inversion. As seen from~\eq{3rdord}, the population-pulsation contribution can
be neglected if $|D_\parallel (\omega_m - \omega_l)| \ll 1$. Typically, the
lasing modes are excited within the gain bandwidth~$\gamma_\perp$ around the
atomic frequency. Then, the above condition reduces to~$\gamma_\perp \gg
\gamma_\parallel$.

Requiring that $\Delta n_\omega (\mbf r) = \Delta n (\mbf r)\, \delta
(\omega)$, we express $\mbf P_\omega$ from~\eq{poleqft} and insert it
in~\eq{inveqft}. If this assumption were actually consistent with the
equation~\eq{inveqft} one would, after carrying out mode-of-the-field
expansion~\eq{lasmod},  end up (in the frequency representation) with terms,
which were also proportional to $\delta(\omega)$. In reality, in addition to
``correct'' terms one would obtain a number of terms proportional to
$\delta$~functions of various combinations of lasing frequencies, which
describe  oscillations of the population. Neglecting this ``oscillatory'' terms
is equivalent to keeping only diagonal contributions $|\mbf E_l|^2$ in the
modal expansion of the $\mbf E^* \cdot \mbf P$ term, quadratic in the field. In
this approximation $\Delta n (\mbf r)$~can be determined selfconsistently and
inserted into~\eq{poleqft} and~\eq{weqactft} to obtain the lasing equations
\begin{eqnarray}
  \fl
  \l\{ \frac \rmd {\rmd t} + \rmi \l[ \Omega_k (\omega_m) - \omega_m \r] \r\}
  a_{km} (t) \nonumber \\
  \fl
  = 2 \pi \nu \frac {d^2} {\hbar \gamma_\perp} D(\omega_m)  \int \rmd \mbf r\,
  \frac {\Delta n_0 (\mbf r)} {\sqrt{\epsilon (\mbf r)}} \frac {\bphi_k^*
  (\omega_m; \mbf r) \cdot \mbf E_m (\mbf r, t)} {1 + \frac {d^2} {\hbar^2
  \gamma_\perp \gamma_\parallel} \sum_l \mrm{Re}\, [D(\omega_l)]\, |\mbf E_l
  (\mbf r, t)|^2}.
\label{allord}
\end{eqnarray}
In contrast to~\eq{3rdord}, the field here is expanded in the quasimodes of
the passive system with the frequencies~$\Omega_k (\omega_m)$, while the linear
mode coupling is included in the right-hand side. These equations represent
generalization of time-independent equations derived in~\cite{ture06},
which are obtained from~\eq{allord} by assuming time independence of the
respective amplitudes.  Equivalently,
equation~\eq{allord} can be derived by treating the polarization term
in~\eq{weqactft} as a source and using the Green function to write down
the solution of this equation as~\cite{ture06}:
\begin{eqnarray}
  \fl
  \mbf E_\omega (\mbf r) &= - \frac {4 \pi \nu^2} {\sqrt{\epsilon (\mbf r)}}
  \int \rmd \mbf r'\, \epsilon^{-1/2} (\mbf r')\, G(\omega; \mbf r, \mbf r')\,
  \mbf P_\omega (\mbf r') \nonumber \\
  \fl
  &= \rmi \frac {4 \pi \nu^2} {\sqrt{\epsilon (\mbf r)}} \frac {d^2} {\hbar
  \gamma_\perp} D(\omega) \int \rmd \mbf r'\, \frac {\Delta n_0 (\mbf r')}
  {\sqrt{\epsilon (\mbf r')}} \frac {G(\omega; \mbf r, \mbf r')\, \mbf E_\omega
  (\mbf r')} {1 + \frac {d^2} {\hbar^2 \gamma_\perp \gamma_\parallel} \sum_l
  \mrm{Re}\, [D(\omega_l)]\, |\mbf E_l (\mbf r, t)|^2}.
\end{eqnarray}
Replacing the Green function with its spectral representation~\eq{GF} in the
rotating-wave approximation and integrating as in~\eq{ak}, we again arrive at
equation~\eq{allord}. In the stationary case this equation is reduced to a
nonlinear eigenmode problem
\begin{eqnarray}
  \fl
  \sum_{k'} T_{kk'} (\omega_m)\, a_{k'm} = p^{-1}\, a_{km},
\label{Teq}\\
  \fl
  T_{kk'} (\omega) = \rmi 2 \pi \nu \frac {d^2} {\hbar \gamma_\perp}
  \frac{D(\omega)} {\omega - \Omega_k (\omega)} \int \rmd \mbf r'\, \frac
  {\delta n_0 (\mbf r')} {\epsilon (\mbf r')} \frac {\bphi_k^* (\omega; \mbf
  r') \cdot \bpsi_{k'} (\omega; \mbf r')} {1 + \frac {d^2} {\hbar^2
  \gamma_\perp \gamma_\parallel} \sum_l \mrm{Re}\, [D(\omega_l)]\, |\mbf E_l
  (\mbf r)|^2}.
\end{eqnarray}
where the unsaturated population inversion $\Delta n_0 (\mbf r) = p\, \delta
n_0 (\mbf r)$ is split into the overall pump strength~$p$ and the pump
profile~$\delta n_0 (\mbf r)$. The field distribution in mode~$m$ is $\mbf E_m
(\mbf r) = \sum_k a_{km} \bpsi_k (\omega_m; \mbf r) / \sqrt{\epsilon (\mbf
r)}$. If the basis of constant-flux modes (section~\ref{sub:cf}) is used, the
field outside of the system can be obtained by continuation.

An algorithm to determine the lasing-mode frequencies~$\omega_m$ and expansion
coefficients~$a_{km}$ as $p$ is increased continuously from zero is described
in~\cite{ture08}. Below the threshold, where all $a_{km} = 0$, one looks for
the eigenvalues of the linear $T (\omega)$. Changing~$\omega$, the eigenvalues
can be made real, one at a time. The largest real eigenvalue $p_1^{-1}$ yields
the threshold pump strength and the corresponding $\omega$ is the lasing
frequency at the threshold. Above the threshold, for $p > p_1$, the pump is
increased in small steps and the solution~$a_{k1} (p)$ for the first mode is
determined iteratively from~\eq{Teq}. The second mode appears when the second
largest eigenvalue $p_2^{-1}$ of $T (\omega)$ linearized ``around'' the first
mode becomes equal to~$p^{-1}$. The procedure is continued to find higher modes.

\section{Examples and properties of multimode random lasers
\label{examp}}

\subsection{Threshold and number-of-modes statistics
\label{modstat}}

Distribution of thresholds and average number of lasing modes as a function of
pump strength was calculated in~\cite{misi98} for an ensemble of weakly open
chaotic cavities. Each cavity was opened via $M$ small holes (diameter $\ll$
wavelength), which together carry $M$ open channels. The passsive-mode decay
rates~$\kappa$ in this system are distributed according to the
$\chi^2$~distribution with $M$ degrees of freedom,
\eqn{
  P_M (y) = \frac {(M/2)^{M/2}} {\Gamma (M/2)} y^{M/2 - 1} \exp \l( - \frac M 2
  y \r), \quad y \equiv \kappa / \langle \kappa \rangle,
}
where $\Gamma (x)$ is the gamma function. The distribution is wide for
small~$M$ and, for $M = 1$, it increases as $y^{-1/2}$ when $y \to 0$. This
property leads to a wide distribution of lasing thresholds, $\Delta n_{0,
\mrm{thr}}$, which behaves as $\Delta n_{0, \mrm{thr}}^{M/2 - 1}$ for
small~$\Delta n_{0, \mrm{thr}}$. The average threshold is much less than the
nominal value $\Delta \wtilde n_0 = \langle \kappa \rangle\, \hbar \gamma_\perp
\epsilon/ 2 \pi \nu d^2$, which is the pumping required to overcome the average
loss at $\omega = \nu$ [cf.~\eq{Ombar}].  An increase of threshold fluctuations
with localization was observed in a one-dimensional disordered
model~\cite{wu07}.

Considering the rate equations~\eq{rateeq} in the stationary regime and
neglecting the population-pulsation term containing~$D_\parallel (\omega_m -
\omega_l)$, one obtains a matrix equation for the intensities $I_m = |\wbar
a_m|^2$ of lasing modes:
\eqn{
\fl
  \sum_l A_{ml}\, I_l = 1 - \frac {\Delta \wtilde n_0} {\Delta n_0} \frac {y_m}
  {\mc L_m}, \quad A_{ml} = \frac {2 d^2 V} {\epsilon \hbar^2 \gamma_\perp
  \gamma_\parallel} \frac {\mc L_l} {\mc L_m} \mrm{Re}\, \l[D(\omega_m)\, B_{ml}
  \frac {\epsilon^2} {V^2 \Delta n_0} \r],
}
where $y_m \equiv \kappa_m / \langle \kappa \rangle$ ($\kappa_m$ is the
decay rate of mode~$m$) and $\mc L_m \equiv \mrm{Re}\, D(\omega_m)$. This
equation must be  complemented by the condition that it has only positive
solutions. The rest of the basis are nonlasing modes and their intensities are
set to zero. It should be noted, however, that the positiveness of lasing
intensity does not, by itself,  guarantee that the found solution is stable,
the fact well known for simple two-mode models~\cite{sarg74}. The stability of
the solutions can only be verified from the time-dependent
equations~\eq{rateeq}, therefore, the estimates of the number of modes based on
time-independent equations cannot, in general, be considered as completely
accurate.

For a weakly open chaotic cavity~\cite{misi98} one can assume that (a)~the
eigenfunctions are almost real and orthogonal and (b)~they can be described as
random Gaussian functions which are uncorrelated for different
modes~\cite{berr77}. If, in addition, one assumes that the background
dielectric constant, $\epsilon (\mrm r)$, and pumping rate, $\Delta n_0(\mrm
r)$, are both uniform, the correlator~\eq{corr} takes the form~of
\eqn{
  B_{ml}\, \epsilon^2/V^2 \Delta n_0 = 1 + 2 \delta_{ml}.
\label{corchao}
}
In this case matrix $A_{ml}$ can be inverted analytically yielding a dependence
of mode intensities on the pump strength. Let us order the modes $m = 1,2,
\ldots$ in the order they are excited as the pump increases. It can be shown
that, in this case, $y_m/\mc L_m$ form an increasing sequence. A threshold
condition for the mode~$m$ results in the equation~\cite{misi98}:
\eqn{
  \l(\frac m 2 + 1 \r) \frac {y_m} {\mc L_m} - \frac 1 2 \sum_{l = 1}^m \frac
  {y_l} {\mc L_l} = \frac {\Delta n_0} {\Delta \wtilde n_0},
\label{Nmchao}
}
which relates the number of excited modes $N_{\mrm m} = m$ and the pump
strength~$\Delta n_0$ in a particular cavity. The ensemble average $\langle
N_{\mrm m} \rangle$ is calculated~\cite{misi98} with the help of the
distribution~$P_M (y)$ for $y \ll 1$ and is found to scale as~$\Delta n_0^{M/(M
+ 2)}$ asymptotically for large pump strength.

In~\cite{hack05} $\langle N_{\mrm m} \rangle$ was computed for nonweakly open
cavities modelled by random matrices. In particular, validity of the
relation~\eq{corchao} was studied numerically for strong coupling to the bath
$\gamma_n = 1$ (section~\ref{RMT}). It was shown that for $\kappa \lesssim
\langle \kappa \rangle$ the assumption of uncorrelated eigenvectors works well,
while for larger~$\kappa$ deviations from Gaussian statistics become stronger.

The power-law asymptotics for $\langle N_{\mrm m} \rangle$ was confirmed by
numerical simulation of decay rates entering~\eq{Nmchao} using random matrices
with $\gamma_n \ll 1$~\cite{zait06}. The standard deviation~$\sigma_{N_{\mrm
m}}$ varied as~$\Delta n_0^{M/2(M + 2)}$. In the case $\gamma_n = 1$, the ratio
$\sigma_{N_{\mrm m}}/\langle N_{\mrm m} \rangle$, but not $\sigma_{N_{\mrm m}}$
and $\langle N_{\mrm m} \rangle$ separately, obeyed a power law with an
exponent that depended on~$M$. The difference in the results for the weak and
strong coupling occurs, because the decay-rate distribution for $\gamma_n \sim
1$ is no longer of $\chi^2$ type~\cite{fyod97}.

In a one-dimensional disordered system~\cite{zait09} assumption~\eq{corchao}
fails completely, indicating a different type of mode competition compared to
chaotic systems. Numerical simulations showed that the number of lasing modes
saturates below the basis size with increasing pump. This effect is related to
the nonmonotonic dependence of mode intensities on the pump strength and
complete disappearance of some modes for pump exceeding certain thresholds.

The mode suppression was also reported in disordered disk lasers studied within
the theory of section~\ref{aot}~\cite{ture08}. Comparing the dependences of
lasing frequencies and intensities on the pump, it was noticed that when two
modes come close together in frequency, one of them can be suppressed. The mode
thresholds and intensities, but not the frequencies, were found to be very
sensitive to the pump spatial profile.

\subsection{Frequency and intensity statistics
\label{mrep}}

It is well known that passive closed chaotic systems without spatial symmetries
have level repulsion, i.e., the probability density for zero frequency spacing
vanishes. A natural question is, how the spacing distribution for lasing modes
in a random laser is connected to the distribution for passive modes in the
underlying system without gain? It can be seen in the following examples that
mode selection and competition normally enhances the repulsion in a laser. When
lasing modes are close to passive modes, this property is rather obvious: even
if two passive modes cross, not both of them will necessarily lase.

Spacing distributions in two-mode chaotic lasers modelled with random matrices
were computed numerically in~\cite{zait07}. Mode repulsion was present both in
the cases of weak and intermediate openness, $\gamma_n \ll 1$ and $\gamma_n =
1$ (section~\ref{RMT}), even though the passive frequencies can cross for
$\gamma_n = 1$. (If two passive modes have the same frequencies, they have
quite different lifetimes due to repulsion in the complex plane. Hence, only
one of the two modes will be lasing for moderate pump strength.) When the gain
bandwidth $\gamma_\perp$ is close to the mean level spacing $\wbar{\Delta
\omega}$ in the passive system, the spacing distribution for the lasing modes
is well described by the Wigner surmise~\cite{haak01}
\eqn{
  P_{\mrm W} (\Delta \omega) = \frac \pi 2 \frac {\Delta \omega}
  {\wbar{\Delta \omega}^2} \exp \l( - \frac \pi 4 \frac {\Delta \omega^2}
  {\wbar{\Delta \omega}^2} \r),
}
derived for passive closed chaotic cavities in the random-matrix theory. Again,
this form works also for $\gamma_n = 1$, when the spacing statistics for
passive modes is closer to Poissonian. This example shows that formal
coincidence between a spacing distribution for lasing modes and the Wigner
surmise does guarantee that the physics behind them is the same.

Mode repulsion with deviations from the Wigner surmise was found numerically in
one-dimensional disordered lasers~\cite{zait09}. There were two reasons for the
repulsion. First, some modes were coupled, because the system was not very long
and the modes could overlap. Second, when two modes were localized (and could
have close frequencies), the mode that was closer to the opening had higher
threshold and was not excited.

When two modes have close frequencies, it may become necessary to take into
account the dependence of mode frequencies on the pump.  As mentioned above, a
correlation between mode repulsion and suppression during the change of the
pump level was observed in a numerical study of two-dimensional disordered
laser~\cite{ture08}.

Spacing distributions were measured in colloidal solutions containing
$\mrm{TiO}_2$ scatterers and a laser dye~\cite{wu08}. The system was in the
weak-scattering regime in the sense that the scattering mean free path was much
longer than the pump excitation cone. The lasing frequencies were more or less
regularly spaced, exhibiting the mode repulsion. The spacing distribution had a
maximum, but could not be fitted well with the Wigner surmise. The average mode
spacing scaled with the dye concentration. (Increasing the concentration
reduced the gain volume, which led to the reduction of the number of modes.)
The spacing fluctuations increased with the scatterer concentration. Some of
the experimental results were supported by numerical simulations in a
one-dimensional disordered system at the threshold. The statistics of lasing
peaks was compared with the statistics of spontaneous-emission spikes that
appear in the background of the emission spectra. The spikes were attributed to
photons created in single spontaneous-emission events and amplified over long
paths~\cite{muju04,muju07}. Coherent feedback is not required for the
appearance of spikes. The spikes' positions in the spectrum were uncorrelated,
which was reflected in the Poisson spacing statistics.

Spectra with almost equally spaced lasing peaks were obtained in $\mrm{TiO}_2$
colloidal solutions with strong reabsorption outside of the pumped
volume~\cite{wu06}. Weak scattering, on the one hand, and reabsorption, on the
other hand, result in an effective cavity being formed by just two scatterers
located at the ends of the excitation cone. (The ``cavity'' has the maximal
possible length, because the gain grows exponentially with the path length,
while the probability of photons leaving the cavity scales as a power of the
length.) The effective cavity is of the Fabry-Perot type, therefore the lasing
peaks are equidistant and the spacing scales inversely with the cone length.

Another system that shows mode repulsion is porous GaP filled with dye
solution~\cite{mole07}. The transport mean free path was of the order of
$\lambda$ and about three times smaller than the pump spot size. The spacing
distribution could be roughly fitted with the Wigner surmise.

The difference between lasing peaks and spontaneous-emission spikes in
$\mrm{TiO}_2$ colloidal solutions with weak reabsorption~\cite{wu08} (see
above) emerges also in the emission intensity statistics. Two
statistical ensembles were considered: (1)~intensities collected from all
wavelengths in some range $I (\lambda)/ \langle I (\lambda) \rangle$,
normalized by the intensity averaged over many shots at given~$\lambda$, and
(2)~peak and spike heights of the functions $I (\lambda)/ \langle I (\lambda)
\rangle$. For both ensembles the probability distribution showed similar
asymptotic behaviour at large intensities: it had a power-law tail above the
lasing threshold and decayed exponentially below the threshold or in the
absence of scatterers (neat dye solution). Numerically computed distribution of
lasing-mode intensities in a one-dimensional disordered laser~\cite{zait09} had
a power-law decay, as well.

It should be mentioned that a power-law asymptotics may also appear in lasers
with incoherent feedback near the threshold~\cite{lepr07}. Thus, caution should
be exercised when using intensity distribution as a test for coherent lasing.

\subsection{Structure of lasing modes}

One of the important recent development, which is relevant not only for random
lasers, but for the entire field of laser physics is the realization of the
fact that so called lasing modes may differ significantly from modes of passive
cavities. It was noted in~\cite{deyc05a} that spatial nonuniformity of the
refractive index and pumping can result in gain-induced linear coupling between
modes of passive cavities, which results in formation of new modes. These ideas
were taken further in self-consistent theory
of~\cite{ture06,ture07,ture08,ture09}, where no a priori assumptions about the
spatial structure of lasing modes were made and they were found from nonlinear
self-consistent equations~\eq{allord}. Calculations
of~\cite{ture06,ture07,ture08,ture09} found significant modifications of the
spatial profile of lasing modes due to nonlinear mode coupling. However, as we
already mentioned, it is not clear if the systems studied
in~\cite{ture06,ture07,ture08,ture09} can be considered as being in diffusive
regime. At the same time, no changes in the spatial structure of a lasing mode
with increased pumping intensity were found in~\cite{vann01,vann07}, where the
studied structures were clearly identified as being either in
localized~\cite{vann01} or diffusive~\cite{vann07} regimes.

Modification of modes due to presence of gain was observed in numerical
simulations of a one-dimensional random laser below or at the
threshold~\cite{wu07,wu08}, but only in the presence of spatially nonuniform
(local) pumping. While modes in a uniformly pumped system were close to passive
modes and their intensity grew exponentially towards the boundary, the
transition to local pumping changed them substantially: they did not grow
exponentially outside of the gain volume, but still were extended over the
whole system. The number of lasing modes under local excitation was found to be
less than the number of passive modes in the same frequency range, but larger
than the number of passive modes in the reduced system defined by the pump
region. In the case of nonuniform pumping the mode modification appears already
in the linear approximation~\cite{deyc05b}, so that the roles of nonlinear
effects in this simulation is unclear.

Structure of lasing modes was also studied in a system with weak scattering and
strong reabsorption~\cite{wu06}. It was found that in such systems the gain
volume surrounded by strongly absorbing medium forms an effective cavity, where
lasing modes are localized. Numerical simulations below and at the threshold in
two dimensions showed that the lasing modes in this case are very close to the
passive modes of the effective cavity.

\section{Conclusions}

Current research on multimode random lasing is moving along several major
directions, such as (i)~extension of conventional laser theories to open and
irregular systems; (ii)~statistical properties of lasing modes;
(iii)~mechanisms of random lasing (quantum and classical localization,
extended modes), to name a few. Up to date a large number of experimental and
numerical observations are collected. However, the systems are hard to access
analytically, as they consist of a number of strongly interacting components
(electromagnetic field, gain medium, scatterers, boundaries) and several
factors (openness, disorder, nonlinearity, noise) are not negligible within a
wide range of parameters.

Recent important developments in the semiclassical multimode theory and
random-matrix theory added to understanding of the properties of lasing modes
in the stationary regime. To enable a direct comparison with experiments,
mostly performed under pulsed-pumping conditions, it would be desirable to
study time-dependent behaviour and relaxation processes. A detailed analysis
of mode stability, hysteresis phenomena and quantum-noise effects in the
stationary regime is also lacking. To address the role of localization in
lasing feedback at an adequate level, numerical simulations of more realistic
(three-dimensional) models might be necessary.

\ack

Financial support for O.Z. was provided by the Deutsche
Forschungs\-gemein\-schaft via the SFB/TR12, while L.D. thanks PSC-CUNY
for partial support of this work.

\section*{References}

\bibliography{laser}

\end{document}